\documentclass{article}

\usepackage[utf8]{inputenc} 
\usepackage[T1]{fontenc}    
\usepackage{hyperref}       
\usepackage{url}            
\usepackage{booktabs}       
\usepackage{amsfonts}       
\usepackage{nicefrac}       
\usepackage{microtype}      
\usepackage{lipsum}
\usepackage{graphicx}
\usepackage{amsmath}
\graphicspath{ {./images/} }

\title{On the stability of equilibria of the physiologically-informed dynamic causal model}

\author{
    Sayan Nag \\

    \texttt{nagsayan112358@gmail.com}
}

\begin{document}
\maketitle
\begin{abstract}
Experimental manipulations perturb the neuronal activity. This phenomenon is manifested in the fMRI response. Dynamic causal model and its variants can model these neuronal responses along with the BOLD responses \cite{friston2003dynamic,stephan2008nonlinear,moran2009dynamic,daunizeau2009variational,havlicek2015physiologically} . Physiologically-informed DCM (P-DCM) \cite{havlicek2015physiologically} gives state-of-the-art results in this aspect. But, P-DCM has more parameters compared to the standard DCM model and the stability of this particular model is still unexplored. In this work, we will try to explore the stability of the P-DCM model and find the ranges of the model parameters which make it stable.
\end{abstract}


\section{Introduction}
Experimental perturbations lead to the fluctuations in the neuronal activity which are known as neuronal responses. These neuronal responses drive the vasoactive signals which in turn drive the blood flow responses. The changes in the blood flow responses lead to changes in the blood volume and deoxyhemoglobin content which together give rise to BOLD signals.

To model this entire flow which occurred in the brain due to some activity and find the causal interactions between the different inter-connected brain regions biophysical models like Dynamic Causal Models (DCMs) have been generated. DCM explains the causality between different regions. The connection between different regions are given in the form of a probabilistic graphical model where the nodes represent the brain regions. The effective connectivity between these regions are parameterized in terms of the coupling parameters. In other words, the effective connectivity between two regions or nodes is expressed in terms of the conditional dependency between those two regions.

The physiologically-informed DCM (P-DCM) is a physiologically realistic state-of-the-art DCM variant which gives better model evidence when compared to other DCM approaches. It introduces a factor in the connection between the excitatory and inhibitory neuronal population. Along with this, there are some modifications in the neurovascular coupling where a feed-forward model has been introduced instead of feedback-based model. It also introduces a dynamic viscoelastic effect in the balloon model. Altogether, it introduces more parameters compared to the vanilla DCM, renders dynamic stability to the system and makes the model more realistic from physiological point of view.

Equilibrium points are not always stable. It is important to analyze the stability of the equilibrium points they play different roles in determining the dynamics of a system. In this work, considering each of the model equations we will look into the parameters and their corresponding ranges which make the pDCM model dynamically stable near respective equilibrium points.

\section{P-DCM Model}

The P-DCM model consists of 3 sub-models, namely, the neuronal model, the neurovascular coupling, the hemodynamic and balloon models and the BOLD model. The BOLD model consists a set of equations which is empirically determined. We will focus on the neuronal model, the neurovascular coupling and the hemodynamic and balloon models which were the main contributions of the P-DCM paper \cite{havlicek2015physiologically}.

\subsection{Neuronal model}
The neuronal model consists of the excitatory and inhibitory states. The dynamics are dictated by the following state-space differential equations:
\begin{equation}
\frac{dx_{E}(t)}{dt} = -\sigma x_{E}(t) - \mu x_{I}(t) + cu(t)
\end{equation}
\begin{equation}
\frac{dx_{I}(t)}{dt} = \lambda x_{E}(t) - \lambda x_{I}(t)
\end{equation}
Here, $\sigma$ represents the self-intrinsic connectivity of the excitatory neuron. $\lambda$ is the inhibitory gain factor. The excitatory-inhibitory mutual connection is given by $\mu$.

\subsection{Neurovascular coupling (NVC)}
Neurovascular coupling refers to the changes in CBF and CBV due to change in the neuronal activity. A feedforward based NVC model has been chosen in P-DCM which is given by the following set of differential equations:

\begin{equation}
\frac{da(t)}{dt} = -\rho a(t) + x_{E}(t)
\end{equation}
\begin{equation}
\frac{df(t)}{dt} = \phi a(t) - \chi (f(t) - 1)
\end{equation}

Here, a(t) is the vasoactive signal. Using the above equations, it transforms the neuronal response $x_{E}(t)$ to the blood flow response which is given by f(t), $\rho$, $\phi$, $\chi$ respectively represents the decay of vasoactive signal, the gain of vasoactive signal and the decay of blood inflow signal.

\subsection{Hemodynamic and Balloon models}

The hemodynamic model dictates the dynamics of the hemodynamic variables called blood volume v(t) and the deoxyhemoglobin content q(t) with respect to the blood inflow f(t) and the blood outflow $f_{out}(v)$.

\begin{equation}
\frac{dv(t)}{dt} = \frac{1}{t_{MTT}}[f(t) - f_{out}(v,t)]
\end{equation}
\begin{equation}
\frac{df(t)}{dt} = \frac{1}{t_{MTT}}[f(t)\frac{E(f)}{E_{0}} - f_{out}(v,t)\frac{q(t)}{v(t)}]
\end{equation}

Here, $t_{MTT}$ is the mean transit time the blood takes to pass the veins. E(f) is the oxygen extraction fraction when the blood inflow is f(t) and $E_{0}$ is the net oxygen extraction at rest. E(f) is given as:

\begin{equation}
E(f) = 1 - (1 - E_{0})^{1/f}
\end{equation}

Considering the visco-elastic effect the balloon model is given as:

\begin{equation}
f_{out}(v,t) = \frac{1}{\tau + t_{MTT}}(t_{MTT} v(t)^{1/\alpha} + \tau f(t))
\end{equation}

Here, $\tau$ is the visco-elastic time constant and $\alpha$ is the Grubb's exponent.

\section{Analysis of P-DCM Model}

Linear Stability Analysis refers to the stability analysis of a system after the linearization of the differential equations at respective equilibrium points are done. Equilibrium refers to the state of a system which does not change. Equilibrium point(s) can be estimated by setting the derivative(s) of the differential equation(s) (describing the system) to zero. 

A Jacobian matrix is formed by the first order partial derivatives of a system of differential equations. The eigen values of a Jacobian matrix can be represented in the form of complex numbers. The complex part of the eigenvalue contributes an oscillatory component, but, the real part determines the stability of the equilibrium point.

An equilibrium point of the differential equation defining a system is considered to be: (a) stable if all the eigenvalues of the Jacobian matrix evaluated at that equilibrium point have negative real parts and (b) unstable if at least one of the eigenvalues of the Jacobian matrix evaluated at that equilibrium point has a positive real part.

In this section we will show the detailed analysis of P-DCM model, especially we will try to investigate the values of the model parameters which make the model dynamically stable near the respective equilibrium points.

\subsection{Neuronal model}

Equations (1) and (2) give the neuronal dynamics of the P-DCM Model. Now, making both of them equal to 0, we will get the equilibrium values of the system. The equilibrium values of $x_E$ and $x_I$ are the same and equals to $cu/(\sigma + \mu)$. Now, we will try to find the nature of stability of the equilibrium point by computing the Jacobian matrix. The Jacobian matrix is represented as:

\begin{equation}
    J_{Neuronal Model} = \begin{bmatrix}
    -\sigma & -\mu\\
    \lambda & -\lambda
    \end{bmatrix}
\end{equation}

The characteristic equation in terms of eigen values (s) is given as:

\begin{equation}
    s^2 + (\sigma + \lambda) s + (\mu \lambda + \sigma \lambda) = 0
\end{equation}

$s_{1}$ and $s_{2}$ are the roots of the above equation given as:

\begin{equation}
    s_{1} = -(\sigma + \lambda) + \sqrt{(\sigma + \lambda)^2 - 4(\mu \lambda + \sigma \lambda)}
\end{equation}

\begin{equation}
    s_{2} = -(\sigma + \lambda) - \sqrt{(\sigma + \lambda)^2 - 4(\mu \lambda + \sigma \lambda)}
\end{equation}

So, for the equilibrium point to be a stable one (and hence the system to be stable),
1. $\sigma + \lambda$ has to be more than 0.
2. $\sqrt{(\sigma + \lambda)^2 - 4(\mu \lambda + \sigma \lambda)}$ can be $\leq 0$, 
3. $\sigma + \lambda > \mod{\sqrt{(\sigma + \lambda)^2 - 4(\mu \lambda + \sigma \lambda)}}$, if $\sqrt{(\sigma + \lambda)^2 - 4(\mu \lambda + \sigma \lambda)} > 0$

By definition each of $\sigma$ and $\lambda$ is positive. So the first condition is satisfied. Considering $\lambda$ > 0,for the third condition, we see that $\mu + \sigma$ has to be greater than 0. By, definition $\mu$ is also positive, so third condition holds true. For the second condition we get:

$$
(\sigma - \lambda)^2 \leq 4\mu\lambda
$$
$$
\lambda - 2\sqrt{\mu\lambda} \leq \sigma \leq \lambda + 2\sqrt{\mu\lambda}
$$

The given ranges for $\mu$ and $\lambda$ in the P-DCM paper \cite{havlicek2015physiologically} are:

$$
0 < \mu < 1.5,
$$
$$
0 < \lambda < 0.3
$$

Now, using these ranges for $\mu$ and $\lambda$ in the above expression showing the permissible range of $\sigma$ and also considering $\sigma$ to be positive (by definition and also to ensure stability), we find:

$$
0 < \sigma < 1.642
$$

In the P-DCM paper, $\sigma$ was made to vary within 0 and 1.5. So as already computed above, the mentioned range of sigma (though it can be extended to 1.64) makes the equilibrium a stable focus and thus makes the system asymptotically stable.

\subsection{Neurovascular Coupling (NVC)}

Equations (3) and (4) give the NVC dynamics of the P-DCM Model. Now, making both of them equal to 0, we will get the equilibrium values of the system. The equilibrium values of $a$ and $f$ are respectively $\frac{x_{E}^{*}}{\rho}$ and $\frac{\phi x_{E}^{*}}{\rho \chi} + 1$. Here, $x_{E}^{*}$ is the steady-state value of $x_{E}$. Now, we will try to find the nature of stability of the equilibrium point by computing the Jacobian matrix. The Jacobian matrix is represented as:

\begin{equation}
    J_{NVC} = \begin{bmatrix}
    -\rho & 0\\
    \phi & -\chi
    \end{bmatrix}
\end{equation}

The characteristic equation in terms of eigen values (s) is given as:

\begin{equation}
    (s + \chi)(s + \rho) = 0
\end{equation}

$s_{1}$ and $s_{2}$ are the roots of the above equation given as:

\begin{equation}
    s_{1} = -\chi
\end{equation}

\begin{equation}
    s_{2} = -\rho
\end{equation}

So, for the equilibrium point to be a stable one (and hence the system to be stable), both $\chi$ and $\rho$ have to be positive. Based on the values mentioned in the P-DCM paper ($\chi = 0.6$ and $\rho = 0.6$), the equilibrium is a stable focus.

\subsection{Hemodynamic and Balloon models}

Putting the values of $E(f)$ and $f_{out}(v,t)$ in equations (5) and (6) and making both of them equal to 0, we will get the equilibrium values of the variables of the system. The equilibrium values of $v$ and $q$ are respectively $v_{eq}$ and $q_{eq}$.

\begin{equation}
    v_{eq} = (f^{*})^{\alpha}
\end{equation}

\begin{equation}
    q_{eq} = \frac{v_{eq} f^{*} E^{*}}{f_{out}^{*} E_{0}}
\end{equation}

Here, $f^{*}$, $f_{out}^{*}$ and $E^{*}$ are respectively the steady-state values of $f$, $f_{out}$ and $E$.

Now, we will try to find the nature of stability of the equilibrium point by computing the Jacobian matrix. The Jacobian matrix is represented as:

\begin{equation}
    J_{H} = \begin{bmatrix}
    -\frac{v_{eq}^{1/\alpha - 1}}{\alpha (t_{MTT} + \tau)} & 0\\
    (\frac{1}{\alpha} - 1)\frac{q_{eq}v_{eq}^{(1/\alpha - 2)}}{t_{MTT} + \tau} + \frac{\tau q_{eq} v_{eq}^{-2}}{t_{MTT}} & -(\frac{v_{eq}^{(1/\alpha - 1)}}{t_{MTT} + \tau} + \frac{\tau f^{*}}{t_{MTT} v_{eq}})
    \end{bmatrix}
\end{equation}

So, the eigen values $s_{1}$ and $s_{2}$ are given as:

\begin{equation}
    s_{1} = -\frac{v_{eq}^{1/\alpha - 1}}{\alpha (t_{MTT} + \tau)}
\end{equation}

\begin{equation}
    s_{2} = -(\frac{v_{eq}^{(1/\alpha - 1)}}{t_{MTT} + \tau} + \frac{\tau f^{*}}{t_{MTT} v_{eq}})
\end{equation}

So, for the equilibrium point to be a stable one (and hence the system to be stable), both $s_{1}$ and $s_{2}$ have to be negative. Based on the values mentioned in the P-DCM paper ($1 < t_{MTT} < 5$ and $0 <\tau < 30$), the equilibrium is a stable focus.

\section{Conclusions}

Linear stability analysis near the equilibrium points for the system of differential equations defining the P-DCM model gives an idea about the ranges of the parameters which make the model dynamically stable. It has been found that the ranges of the model parameters defined in the actual P-DCM paper \cite{havlicek2015physiologically} confer dynamical stability on the equilibrium points of the model differential equations.

\bibliographystyle{unsrt}
\bibliography{references}  

\begin{thebibliography}{1}

\bibitem{friston2003dynamic}
Karl~J Friston, Lee Harrison, and Will Penny.
\newblock Dynamic causal modelling.
\newblock {\em Neuroimage}, 19(4):1273--1302, 2003.

\bibitem{stephan2008nonlinear}
Klaas~Enno Stephan, Lars Kasper, Lee~M Harrison, Jean Daunizeau, Hanneke~EM den
  Ouden, Michael Breakspear, and Karl~J Friston.
\newblock Nonlinear dynamic causal models for fmri.
\newblock {\em Neuroimage}, 42(2):649--662, 2008.

\bibitem{moran2009dynamic}
Rosalyn~J Moran, Klaas~E Stephan, T~Seidenbecher, H-C Pape, Raymond~J Dolan,
  and Karl~J Friston.
\newblock Dynamic causal models of steady-state responses.
\newblock {\em Neuroimage}, 44(3):796--811, 2009.

\bibitem{daunizeau2009variational}
Jean Daunizeau, Karl~J Friston, and Stefan~J Kiebel.
\newblock Variational bayesian identification and prediction of stochastic
  nonlinear dynamic causal models.
\newblock {\em Physica D: nonlinear phenomena}, 238(21):2089--2118, 2009.

\bibitem{havlicek2015physiologically}
Martin Havlicek, Alard Roebroeck, Karl Friston, Anna Gardumi, Dimo Ivanov, and
  Kamil Uludag.
\newblock Physiologically informed dynamic causal modeling of fmri data.
\newblock {\em Neuroimage}, 122:355--372, 2015.

\end{thebibliography}

\end{document}